\documentstyle[12pt,epsf]{article}

\advance \topmargin by -\headheight
\advance \topmargin by -\headsep

\evensidemargin \oddsidemargin
\def\ie{{\em i.e.}}

\def\ie{\hbox{\it i.e.}}

\def\CC{{\mathchoice
{\rm C\mkern-8mu\vrule height1.45ex depth-.05ex 
width.05em\mkern9mu\kern-.05em}
{\rm C\mkern-8mu\vrule height1.45ex depth-.05ex 
width.05em\mkern9mu\kern-.05em}
{\rm C\mkern-8mu\vrule height1ex depth-.07ex 
width.035em\mkern9mu\kern-.035em}
{\rm C\mkern-8mu\vrule height.65ex depth-.1ex 
width.025em\mkern8mu\kern-.025em}}}

\def\RR{{\rm I\kern-1.6pt {\rm R}}}

\def\ZZ{{\rm Z}\kern-3.8pt {\rm Z} \kern2pt}

\def\np{Nucl. Phys.}
\def\pl{Phys. Lett.}
\def\prl{Phys. Rev. Lett.}
\def\pr{Phys. Rev.}

\def\mpl{Mod. Phys. Lett.}

\def\sjnp{Sov. J. Nucl. Phys.}

\def\rmp{Rev. Mod. Phys.}

\def\jhep{J. High Energy Phys.}

\newcommand{\beq}{\begin{equation}}
\newcommand{\eeq}{\end{equation}}
\newcommand{\rc}{\nonumber\\}
\newcommand{\bear}{\begin{eqnarray}}
\newcommand{\eear}{\end{eqnarray}}
\newcommand{\ba}{\begin{array}}
\newcommand{\ea}{\end{array}}

\newfont{\namefont}{cmr10}
\newfont{\addfont}{cmti7 scaled 1440}
\newfont{\boldmathfont}{cmbx10}
\newfont{\headfontb}{cmbx10 scaled 1728}
\begin{document}
\begin{titlepage}

\begin{center} \Large \bf World-volume Solitons of the D3-brane in
the Background of  $(p,q)$ Five-branes

\end{center}

\vskip 0.3truein
\begin{center} 
P. M. Llatas
\footnote{e-mail:llatas@fpaxp1.usc.es}, 
A.V. Ramallo
\footnote{e-mail:alfonso@fpaxp1.usc.es}
and 
J. M. S\'anchez de Santos 
\footnote{e-mail:santos@gaes.usc.es}

\vspace{0.3in}

Departamento de F\'\i sica de
Part\'\i culas, \\ Universidad de Santiago\\
E-15706 Santiago de Compostela, Spain. 
\vspace{0.3in}

\end{center}
\vskip 1truein

\begin{center}
\bf ABSTRACT
\end{center} 

We analyze the world-volume solitons of a D3-brane probe in the
background of parallel $(p,q)$ five-branes. The D3-brane is
embedded along the directions transverse  to the five-branes of the
background. By using the S-duality invariance of the D3-brane, we find
a first-order differential equation whose solutions saturate an energy
bound. The $SO(3)$ invariant solutions of this equation are found
analytically. They represent world-volume solitons which can be
interpreted as formed by parallel $(-q,p)$ strings emanating from the
D3-brane world-volume. It is shown that these configurations are $1/4$
supersymmetric  and provide a world-volume realization of the
Hanany-Witten effect.

\vskip4.5truecm
\leftline{US-FT-25/99 \hfill December 1999}
\leftline{hep-th/9912177}
\smallskip
\end{titlepage}
\setcounter{footnote}{0}

\setcounter{equation}{0}
\section{Introduction}
\medskip

The study of the solitons of the world-volume theories which describe
the dynamics of branes has provided a lot of  information about the
different aspects of brane physics and of their interrelations
\cite{review}. An interesting aspect of these soliton solutions is  that
they have the spacetime interpretation of branes ending on the
world-volume. In fact, different supergravity solutions, which
represent spacetime configurations of intersecting branes, have been
realized as solitons on the world-volume \cite{CM, Gibbons, HLW}. These
results have allowed to perform several consistency checks of the
different approaches to describe  branes in string theory  and of its
connections to Yang-Mills theories. 

Of particular interest are the world-volume solitons of brane probes
which move in the supergravity background created by another brane. In
some of these cases the background acts as a source for the
world-volume gauge field and, thus, induces a soliton on the
world-volume of the brane probe. The resulting configuration is a
triple intersection of the background, probe and soliton branes. 

An example of  this last type of systems is the baryon vertex
\cite{Wittenbaryon}, in which a D5-brane moves in $AdS_5\times S^5$,
which is the near-horizon geometry of a stack of parallel  D3-branes. A
BPS equation for the embedding of the D3-brane in $AdS_5\times S^5$ was
proposed in ref. \cite{Imamura}. This equation was analyzed in ref.
\cite{CGS1},  where it was generalized to the full geometry of the
D3-brane background. In ref. \cite{Craps}  the BPS condition was
reobtained as a saturation condition of an energy bound.
These results were generalized in ref. \cite{baryon}, where the system
of a D$(8-p)$-brane in the background of a D$p$-brane for $p\le 6$ was
studied and the corresponding BPS condition was obtained and integrated
analytically. Moreover, in ref. \cite{kappa} it was shown that the BPS
condition obtained from the energy argument is precisely the one needed
to preserve $1/4$ supersymmetry. In all these systems the world-volume
soliton is a spike which can be interpreted as a bundle of fundamental
strings emanating from the brane probe.

In this paper we extend the results of refs.
\cite{Imamura}-\cite{kappa} to the case of a D3-brane probe in the
background of a stack of parallel
$(p,q)$ five-branes (the case of a D3-brane in the background of 
Neveu-Schwarz five-branes was studied in ref. \cite{Joan}). The
$(p,q)$ five-brane is a magnetically charged object under both the
Neveu-Schwarz (NS) and Ramond (R) three-form field strengths of the
type IIB supergravity,
\ie\ is the magnetic analogue of the $(p,q)$ strings studied by 
Schwarz \cite{Schwarz}. It can be regarded as a bound state of  $p$
NS5-branes and $q$ D5-branes. The complete form of this background was
obtained in ref. \cite{LuRoy} by using the $SL(2,\ZZ)$ S-duality
symmetry of the type IIB supergravity.

The S-duality symmetry will be very important in our approach. Indeed,
it is well-known \cite{review} that the D3-brane is invariant under this
symmetry. At the level of the world-volume action, this invariance is
reflected by the fact that one can convert the problem of a D3-brane in
a  $(p,q)$ five-brane background into the problem of a D3-brane in a
D5-brane background. A BPS condition for the latter was found in ref.
\cite{baryon} and, thus, by inverting the S-duality transformation, 
one can obtain a BPS condition for the $(p,q)$ five-brane background.
This condition can be integrated analytically and we will show that the
corresponding world-volume soliton can be interpreted as a bundle of
$(-q, p)$ strings. We will also prove that this configuration is $1/4$
supersymmetric. 

This paper is organized as follows. In section 2 we will describe the 
$(p,q)$ five-brane background. In section 3 we will analyze the action
and equations of motion of the D3-brane probe and, in particular we
will put its energy functional in a form in which the
invariance under S-duality is manifest. A BPS condition minimizing the
energy will be found in section 4, where we will also obtain the
explicit form of the world-volume solitons. The analysis of the
supersymmetries preserved by our solution will be the subject of
section 5. In this section we will rederive the BPS condition  as the
one to be imposed if $1/4$ supersymmetry is required. Finally, in
section 6 we summarize our results and discuss some possible
generalizations of our work.

\setcounter{equation}{0}
\section{The $(p,q)$ five-brane background}
\medskip

The massless bosonic fields of the type IIB superstring theory in the
NSNS sector are the graviton $G_{\mu\nu}$, the dilaton $\phi$ and an
antisymmetric Kalb-Ramond field $B_{\mu\nu}$. In the RR sector the 
type IIB theory has an scalar $\chi$, an antisymmetric tensor field
$C_{\mu\nu}$ and a four-form gauge field whose field strength is a
self-dual five-form. When this five-form field strength is zero, the
equations of motion of the type  IIB supergravity \cite{IIB} can be
derived from an action whose bosonic part, written  in the Einstein
frame,  is given by:

\bear
S_{IIB}\,&=&\,\int\,d^{10}x\,\sqrt{-G}\,
\Biggl[\,R\,-{1\over 2}\,\nabla_{\mu}\,\phi\,\nabla^{\mu}\phi\,-\,
{1\over 2}\,e^{2\phi}\,\nabla_{\mu}\,\chi\,\nabla^{\mu}\chi\,-\rc\rc
&&\,-\,{1\over 12}\,\Bigl[\,e^{-\phi}\,
H_{\mu\nu\lambda}\,H^{\mu\nu\lambda}\,
\,+\,e^{\phi}\,
\Bigl(\,G_{\mu\nu\lambda}\,-\chi\,H_{\mu\nu\lambda}\,\Bigr)
\,\Bigl(\,G^{\mu\nu\lambda}\,-\chi\,H^{\mu\nu\lambda}\,\Bigr)
\Bigr]\,\,\Biggr]\,\,.\rc
\label{uno}
\eear
In this action $H$ is the field strength of the NSNS $B$ field:
\beq
H\,=\,dB\,,
\label{dos}
\eeq
whereas $G$ is given by:
\beq
G\,=\,dC\,.
\label{tres}
\eeq
The action (\ref{uno}),  and the equations of motion derived from it,
are invariant under an $SL(2,\RR)$ group of transformations.  This is
the so-called S-duality of the type IIB effective action, which is a
non-perturbative symmetry of the theory.   The NSNS and RR gauge
fields transform as a doublet of this $SL(2,\RR)$ symmetry, 
whereas the dilaton  and RR scalar are mixed under
S-duality. At the quantum level,  only the discrete 
subgroup $SL(2,\ZZ)$ of  $SL(2,\RR)$ survives. 
Actually, one can generate a multiplet of
solutions by acting with the $SL(2,\ZZ)$ symmetry on a particular
solution. This is the method used by Schwarz 
\cite{Schwarz} to construct new string-like solutions of the type IIB
string theory starting from the fundamental string, {\it i.e.} from a
solution electrically charged under the NSNS gauge field. One gets in
this way a series of extremal solutions labeled by two integers $p$ and
$q$, which are usually referred to as $(p,q)$ strings. 

One can apply the same procedure to the magnetically charged NSNS
solution, which is nothing but the NS five-brane (NS5). This
construction has been undertaken by Lu and Roy in ref. \cite{LuRoy} and
as a result one gets an infinite family of five-branes of the type IIB
effective action. Each member of this family is labeled by two
integers $p$ and $q$, which represent, respectively, the magnetic
charges under the NSNS and RR gauge fields. The $(p,q)=(1,0)$ solution
is the NS5-brane one starts with, the $(p,q)=(0,1)$ is the
D5-brane and a general $(p,q)$ five-brane can be regarded as an
extremal  bound state of $p$ NS5-branes and $q$ D5-branes. Since this is
the type IIB background we are interested in, let us describe it in
detail following ref. \cite{LuRoy}. 

The metric for an stack of $N$ parallel  $(p,q)$ five-branes is given
by:
\beq
ds^2\,=\,\Bigl[\,H_{(p,q)}(r)\,\Bigr]^{-{1\over 4}}\,\,
(\,-dt^2\,+\,dx_{\parallel}^2\,)\,+\,
\Bigl[\,H_{(p,q)}(r)\,\Bigr]^{{3\over 4}}\,\,
(\,dr^2\,+\,r^2\,d\Omega_{3}^2\,)\,\,,
\label{cuatro}
\eeq
where $x_{\parallel}^i$ ($i=1,\cdots,5$) are the coordinates along
which the five-brane is extended and we have used spherical
coordinates to parametrize the directions transverse to the five-brane.
($d\Omega_{3}^2$ represents the metric of the unit three sphere). In
the metric written above,  $H_{(p,q)}(r)$ is an harmonic function:
\beq
H_{(p,q)}(r)\,=\,1\,+\,{R^2_{(p,q)}\over r^2}\,\,,
\label{cinco}
\eeq
where the ``radius" $R_{(p,q)}$ is given by:
\beq
R^2_{(p,q)}\,=\,N\,\Bigl[\,\rho_{(p,q)}\,\Bigr]^{{1\over 2}}
\,\alpha'\,\,.
\label{seis}
\eeq
In eq. (\ref{seis}) $\alpha'$ is the Regge slope and $\rho_{(p,q)}$ can
be put in terms of the string coupling constant $g_{s}$ and of the
asymptotic value of the RR scalar $\chi_{0}$ as follows:
\beq
\rho_{(p,q)}\,=\,g_{s}^{-1}\,p^2\,+\,(\,q\,+\,p\chi_{0}\,)^2\,g_s\,\,.
\label{siete}
\eeq
Let us recall that the string coupling constant $g_{s}$ is related to
the asymptotic value of the dilaton $\phi_0=\phi(r=\infty)$ by means of
the expression:
\beq
g_s\,=\,e^{\phi_0}\,\,.
\label{ocho}
\eeq
The dilaton field  for the stack of $(p,q)$ five-branes  at an
arbitrary value of $r$ is given by:
\beq
e^{-\phi}\,=\,
{\rho_{(p,q)}\,e^{-\phi_0}\over
p^2\,\,e^{-\phi_0}\,\Bigl[\,H_{(p,q)}(r)\,\Bigr]^{{1\over 2}}
\,+\,(\,q\,+\,p\chi_{0}\,)^2\,e^{\phi_0}\,\,
\Bigl[\,H_{(p,q)}(r)\,\Bigr]^{-{1\over 2}}}\,\,,
\label{nueve}
\eeq
whereas the RR scalar takes the form:
\beq
\chi\,=\,{\rho_{(p,q)}\,\chi_0\,+\,
pq\,e^{-\phi_0}\,\Bigl(1\,-\,H_{(p,q)}(r)\,\Bigr)\over
p^2\,\,e^{-\phi_0}\,H_{(p,q)}(r)
\,+\,(\,q\,+\,p\chi_{0}\,)^2\,e^{\phi_0}}\,\,.
\label{diez}
\eeq
In order to determine completely the solution, it remains to give the
values of the NSNS and RR gauge fields. Their field strengths $H$ and
$G$ can be written\footnote{Notice that the action of type IIB supergravity written in
ref. \cite{LuRoy} differs from the one above in the change $G
\rightarrow-G$. Therefore, in order to write the $(p,q)$ five-brane
solution with our conventions, this same change must be done in the
solution of ref. \cite{LuRoy}.} in terms of the volume element
$\epsilon_3$ of the three dimensional sphere as follows:
\beq
H\,=\,2\,p\,N \,\alpha'\,\epsilon_3\,,
\,\,\,\,\,\,\,\,\,\,\,\,\,\,\,\,\,\,\,\,\,\,\,\,
G\,=\,-2\,q\,N \,\alpha'\,\epsilon_3\,.
\label{once}
\eeq
From the values of the field strengths displayed in eq. (\ref{once})
it is clear that the object which generates the metric (\ref{cuatro})
and the values of the dilaton and RR scalar of
eqs. (\ref{nueve}) and (\ref{diez}) is magnetically charged under the
NSNS and RR gauge fields, {\it i.e.} extended along the direction of
the potentials of the Hodge duals of
$H$ and $G$. Moreover, the $(p,q)$ five-brane solution is extremal in
the sense that preserves $1/2$ of the supersymmetries of the type IIB
supergravity. The Killing spinors corresponding to this solution will
be given in section 5.

\setcounter{equation}{0}
\section{The D3-brane probe}
\medskip

Let us now consider a D3-brane probe embedded along the transverse
directions of the stack of $(p ,q)$ five-branes described in sect. 2.
The action of  this D3-brane probe is the sum of the
Dirac-Born-Infeld term and the Wess-Zumino term:
\beq
S\,=\,S_{DBI}\,+\,S_{WZ}\,\,,
\label{doce}
\eeq
where the expressions of $S_{DBI}$ and $S_{WZ}$ in the Einstein frame
are given by:
\bear
S_{DBI}\,&=&\,-\,T_{3}\,\int\,d^4\xi\,
\sqrt{-{\rm det}\,(\,g\,+\,e^{-{\phi\over 2}}\,
{\cal F}\,)}\,\,,\rc\rc
S_{WZ}\,&=&T_{3}\,\int\,d^4\xi\,\,
\Bigl[\,{\cal F}\,\wedge\,C\,+\,
{1\over 2}\,\,\chi\,{\cal F}\,\wedge{\cal F}\,
\Bigr]\,\,.
\label{trece}
\eear
In eq. (\ref{trece}) $g$ is the induced world-volume metric, $C$ is
the pullback of the RR gauge field and ${\cal F}$ is 
\beq
{\cal F}\,=\,dA\,-\,B\,=\,F\,-\,B\,\,,
\label{catorce}
\eeq
where $A$ is a $U(1)$ world-volume gauge field, $F$ is its field
strength and $B$ is the pullback of the Kalb-Ramond field of the NS
sector of the superstring. Notice that in $S_{WZ}$ the field $C$ acts
as a source for the world-volume electric charge, whereas the RR scalar
$\chi$ couples to the instanton number density of ${\cal F}$. The
coefficient
$T_{3}$ multiplying the two terms of the action is the tension of the
D3-brane which, in terms of the Regge slope $\alpha'$, can be written
as:
\beq
T_{3}\,=\,{1\over 8\pi^3\,(\,\alpha'\,)^2}\,\,.
\eeq
Let us parametrize the transverse three-sphere by means of three angles
$\theta^1$, $\theta^2$ and $\theta^3$, where $\theta^3\equiv\theta$ is
the polar angle ($0\le\theta\le\pi$). The line element $d\Omega_3^2$
in these coordinates is given by:
\beq
d\Omega_3^2\,=\,(\,d\theta)^2\,+\,(\,\sin\theta\,)^2\,
[\,(\,d\theta^2\,)^2\,+\,(\,\sin\theta^2\,)^2\,
(\,d\theta^1\,)^2\,]\,\,.
\label{dseis}
\eeq
Let us denote by $\bar g$ the determinant of the metric (\ref{dseis}).
It is clear that:
\beq
\sqrt{\bar g} \,=\,(\,\sin\theta\,)^2\,\sin\theta^2\,\,,
\label{dsiete}
\eeq
and, therefore, the volume element $\epsilon_3$ can be written in
these coordinates as:
\beq
\epsilon_3\,=\,\sqrt{\bar g} \,\,\,d\theta^1\,
\wedge\,d\theta^2\,\wedge\,d\theta\,\,.
\label{docho}
\eeq
We shall take the angles $\theta^i$, together with the time variable
$t$ as world-volume coordinates $\xi^{\alpha}$, {\it i.e.} 
$\xi^{\alpha}\,=\,(\,t\,,\,\theta^1\,,\,\theta^2\,,\,
\theta^3\,)$. Actually, we are going to consider static ({\it i.e.}
time independent) configurations of the D3-brane  which
correspond to a radial deformation of the  probe. Accordingly, the
only ``active scalar" will be the coordinate $r$, which we will take
to depend only on the polar angle $\theta$. These configurations
are invariant under an $SO(3)$ symmetry, which is the
maximal amount of symmetry that a radial deformation can have. As was
mentioned above, the first term in $S_{WZ}$ in eq. (\ref{trece})
couples the world-volume gauge field  to the RR background potential 
$C$. Actually, by using the expression (\ref{catorce}) of ${\cal F}$
and integrating by parts, one can easily verify that this term
of $S_{WZ}$ couples the time component $A_0$ of the $U(1)$ world-volume
field to the RR field strength $G$. 
Because of this coupling, we cannot take
$A_0$ to vanish. In what follows we shall assume that $A_0$ depends
only  on the polar angle $\theta$, which again corresponds to the
maximally symmetric solution. On the other hand, the NS field $B$
couples to the brane probe through ${\cal F}$ (see eq.
(\ref{catorce})). Moreover, it is possible to choose a gauge in which
the only non-vanishing component of $B$ is  $B_{\theta^1\,\theta^2}$.
Indeed, one can take
$B$ as:
\beq
B\,=\,2p\,N\,\alpha'\,\sin\theta^2\,b_{\nu}(\theta)\,
d\theta^1\wedge d\theta^2\,\,,
\label{dnueve}
\eeq
where $b_{\nu}(\theta)$ must verify\footnote{The label $\nu$ of the
function  $b_{\nu}(\theta)$ represents the constant of integration of
equation (\ref{veinte}). The explicit expression of $b_{\nu}(\theta)$
will be given in section 4. }:
\beq
{d\over d\theta}\,b_{\nu}(\theta)\,=\,(\,\sin\theta)^2\,\,.
\label{veinte}
\eeq
In what follows we shall adopt the gauge (\ref{dnueve}) and,
accordingly, we shall assume that the only non-vanishing components of 
${\cal F}$ are ${\cal F}_{0\theta}\,=\,F_{0\theta}$  and 
${\cal F}_{\theta^1\theta^2}\,\equiv\,{\cal F}_{12}$. The action $S$
for these configurations is:
\bear
&&S\,=\,T_3\,T\,\int\,d^3\theta\,\Biggl\{
-\sqrt{r^4\,\Bigl[\,H_{(p,q)}(r)\,\Bigr]^{{3\over 2}}\,\bar g\,+\,
e^{-\phi}\,\,{\cal F}_{12}^2}\,\,\times\rc\rc
&&\times\,\,
\sqrt{\Bigl[\,H_{(p,q)}(r)\,\Bigr]^{{1\over 2}}\,
(\,r^2\,+\,r^{\,'\,2}\,)\,-\,
e^{-\phi}\,\, F_{0\theta}^2}\,-\,
2qN\alpha'\,\sqrt{\bar g}\,A_0\,+\,
\chi\,{\cal F}_{12}\,F_{0\theta}\,\Biggr\}\,\,,
\label{vuno}
\eear
where the prime denotes derivative with respect to the polar angle
$\theta$ and $T\,=\,\int dt$. Let us now define the quantity:
\beq
\Pi\,\equiv\,{1\over T_3\,T}\,\,
{\partial S\over \partial F_{0\theta}}\,\,.
\label{vdos}
\eeq
By using the form of $S$ given in eq. (\ref{vuno}), one can find the
expression of $\Pi$ in terms of $F_{0\theta}$ and ${\cal F}_{12}$,
namely:
\beq
\Pi\,=\,
{\sqrt{r^4\,\Bigl[\,H_{(p,q)}(r)\,\Bigr]^{{3\over 2}}\,\bar g\,+\,
e^{-\phi}\,\,{\cal F}_{12}^2}\over
\sqrt{\Bigl[\,H_{(p,q)}(r)\,\Bigr]^{{1\over 2}}\,
(\,r^2\,+\,r^{\,'\,2}\,)\,-\,
e^{-\phi}\,\, F_{0\theta}^2}}\,\,e^{-\phi}\,F_{0\theta}\,+\,
\chi\,{\cal F}_{12}\,\,.
\label{vtres}
\eeq
Moreover, the Euler-Lagrange equation for $A_0$ can be written as the
following   equation for $\Pi$:
\beq
\partial_{\theta}\,\Pi\,=\,2qN\alpha'\,\sqrt{\bar g}\,\,,
\label{vcuatro}
\eeq
which is nothing but the Gauss law constraint for the world-volume
gauge field. Notice that the right-hand side of eq. (\ref{vcuatro})
only depends on the angles and, thus, the Gauss law can be integrated
and, as a result, one can obtain  $\Pi$ as a function of the $\theta$'s.
Similarly, by computing the exterior derivative of ${\cal
F}\,=\,dA\,-\,B$, we get the Bianchi identity for ${\cal F}$:
\beq
d\,{\cal F}\,=\,-H\,\,,
\label{vcinco}
\eeq
which implies that ${\cal F}_{0\theta}$ is independent of the angles
$\theta^1$ and $\theta^2$ (in agreement with our assumption that $A_0$
only depends on $\theta$) and the following differential equation for
${\cal F}_{12}$:
\beq
\partial_{\theta}\, {\cal F}_{12}\,=\,-2pN\alpha'\,\sqrt{\bar g}\,\,.
\label{vseis}
\eeq
The similarity of eqs. (\ref{vcuatro}) and (\ref{vseis}) is manifest.
It is clear that eq. (\ref{vseis}) could also be integrated and one
could obtain ${\cal F}_{12}$ as a function of the angles $\theta$. We
are going to postpone the integration of these equations  until the next
section. However we shall proceed as if $\Pi$ and ${\cal F}_{12}$ were
known functions and we will try to put the action in terms of them. It
is clear from eq. (\ref{vuno}) that one must eliminate $F_{0\theta}$ 
and $A_0$ in favor of $\Pi$ and ${\cal F}_{12}$. This can be done
by inverting the relation between $\Pi$ and $F_{0\theta}$. The result
of this inversion is:
\beq
F_{0\theta}\,=\,
{\sqrt{\Bigl[\,H_{(p,q)}(r)\,\Bigr]^{{1\over 2}}\,
(\,r^2\,+\,r^{\,'\,2}\,)}\over
{\sqrt{r^4\,\Bigl[\,H_{(p,q)}(r)\,\Bigr]^{{3\over 2}}\,\bar g\,+\,
e^{-\phi}\,\,{\cal F}_{12}^2\,+\,
e^{\phi}\,(\,\Pi\,-\,\chi\,{\cal F}_{12}\,)^2 }}}\,\,\,
e^{\phi}\,(\,\Pi\,-\,\chi\,{\cal F}_{12}\,)\,\,.
\label{vsiete}
\eeq
Moreover, by using the Gauss law (\ref{vcuatro}) and integrating by
parts, we can rewrite the WZ term of the action as:
\beq
S_{WZ}\,=\,-T_3\,T\int\,d^3\theta\,
(\,\Pi\,-\,\chi\,{\cal F}_{12}\,)\,F_{0\theta}\,\,.
\label{vocho}
\eeq
By means of eq. (\ref{vsiete}),  and after representing the WZ term as
in eq. (\ref{vocho}), one can write the total action $S$ as:
\beq
S\,=\,-T\,U\,\,,
\label{vnueve}
\eeq
where $U$ is given by:
\bear
&&U\,=\,T_3\,\int\,d^3\theta\,
\sqrt{r^2\,+\,r^{\,'\,2}}\,\,\times\rc\rc
&&\times\,\,
\sqrt{[\,\Sigma_{(p,q)}\,]^2\,+\,
\Bigl[\,H_{(p,q)}(r)\,\Bigr]^{{1\over 2}}\,\,
\Bigl(\,e^{-\phi}\,{\cal F}_{12}^2\,+\,
e^{\phi}\,(\,\Pi\,-\,\chi\,{\cal F}_{12}\,)^2\,\Bigr)}\,\,.
\label{treinta}
\eear
In eq. (\ref{treinta}) $\Sigma_{(p,q)}$ denotes: 
\beq
\Sigma_{(p,q)}\,\equiv\,r^2\,H_{(p,q)}(r)\,\sqrt{\bar g}\,\,.
\label{tuno}
\eeq
It is clear from eq. (\ref{vnueve}) that $S$ and $U$ give rise to the
same Euler-Lagrange equations. In fact, one can regard $U$ as the
result of performing a Legendre transformation to $S$ and, thus, $U$
can be considered as an energy functional for the D3-brane probe in the
$(p,q)$ five-brane geometry. The equations of motion derived from $U$
are simply the conditions required to have  minimal energy
configurations.

An important point, which we shall exploit in the next section, is
that $U$ can be written in a way which makes manifest its $SL(2,\RR)$
invariance. In order to recast $U$ as a function explicitly invariant
under $S$-duality, let us define the matrix:
\beq
{\cal M}\,=\,
\pmatrix{\chi^2\,+\,e^{-2\phi}&\chi\cr\cr
         \chi&1}\,\,e^{\phi}\,\,,
\label{tdos}
\eeq
and the vector
\beq
{\cal D}\,=\,\pmatrix{-{\cal F}_{12}\cr\cr\Pi}\,\,.
\label{ttres}
\eeq
By inspecting eq. (\ref{treinta}) and the definitions 
(\ref{tdos}) and (\ref{ttres}), one immediately verifies that $U$ can
be put as:
\beq
U\,=\,T_3\,\int\,d^3\theta\,
\sqrt{r^2\,+\,r^{\,'\,2}}\,\,
\sqrt{\Sigma_{(p,q)}^2\,+\,\Bigl[\,H_{(p,q)}(r)\,\Bigr]^{{1\over 2}}\,\,
{\cal D}^T\,{\cal M}\,{\cal D}}\,\,.
\label{tcuatro}
\eeq
Let $\Lambda$ be an arbitrary $2\times 2$  $SL(2,\RR)$ matrix of the
form:
\beq
\Lambda\,=\,\pmatrix{a&b\cr c&d}
\,\,,\,\,\,\,\,\,\,\,\,\,
ad-bc\,=\,1\,\,,
\label{tcinco}
\eeq
where $a$, $b$, $c$ and $d$ are constant real numbers. It is now
obvious from eq. (\ref{tcuatro}) that $U$ is invariant under the
following simultaneous transformation of ${\cal M}$ and ${\cal D}$:
\beq
{\cal M}\,\rightarrow\,\Lambda\,{\cal M}\,\Lambda^T\,\,,
\,\,\,\,\,\,\,\,\,\,\,\,\,\,\,\,\,
{\cal D}\,\rightarrow\,(\,\Lambda^{-1}\,)^T\,{\cal D}\,\,.
\label{tseis}
\eeq
Notice that the matrix ${\cal M}$ encodes the information of the
scalar fields of the background. Actually, this same matrix ${\cal M}$
appears when the background action (\ref{uno}) is written in an
$SL(2,\RR)$-invariant form \cite{review}. It is well-known that if one
assembles the RR scalar $\chi$ and the dilaton $\phi$ into a complex
scalar
$\lambda$ as:
\beq
\lambda\,=\,\chi\,+\,i\,e^{-\phi}\,\,,
\label{tsiete}
\eeq
then, the transformation law of  ${\cal M}$ (see eq. (\ref{tseis}))
is equivalent to a M\"obius bilinear transformation of the complex
scalar
$\lambda$:
\beq
\lambda\,\rightarrow\,{a\lambda+b\over c\lambda+d}\,\,.
\label{tocho}
\eeq
It follows from eq. (\ref{tseis}) that the $SL(2,\RR)$ transformation
of the background must be accompanied by a transformation of ${\cal
D}$, which can be regarded as an electric-magnetic duality rotation of
the world-volume gauge fields. Notice that under this symmetry the
Gauss law and the Bianchi identity are mixed. This connection between
world-volume electric-magnetic symmetry and $S$-duality of the
background is well-known \cite{dual}-\cite{coset}. In the next
section we shall use it to map our problem to a system in which the
D3-brane is in the background of a D5-brane, \ie\ we will reduce our
problem for arbitrary values of the integers $p$ and $q$ to the case in
which $(p,q)=(0,1)$. This D5-D3 system was studied in ref.
\cite{baryon} and its equation of motion was integrated exactly by
means of a BPS first-order condition. In section 4 we will be able to
translate this BPS condition of the D5-D3 system into a condition for
the $(p,q)$ five-brane case and, as a consequence, we will find exact
solutions for the embedding of the D3-brane in the general $(p,q)$
background.

\setcounter{equation}{0}
\section{The BPS equation}
\medskip

In this section we will find a first-order (BPS) differential equation
\cite{BPS} whose solutions minimize the energy functional of the
D3-brane \cite{GGT}. Our strategy to find this BPS equation will be the
following. First of all we shall  perform an $S$-duality transformation
which converts the background of eqs. (\ref{cuatro}), (\ref{nueve}),
(\ref{diez}) and (\ref{once}) into a solution of IIB supergravity with
line element given by eq. (\ref{cuatro}) in which only the dilaton
$\phi$ and the RR field strength $G$ are non-vanishing. The problem of
minimizing the energy in the dual variables has been solved in ref.
\cite{baryon}. After transforming back to our original fields, we will
get the minimal energy configurations of our system. 

The form of the $SL(2,\RR)$ matrix with the properties required for
our purposes is not difficult to find. Actually, the $(p,q)$
five-brane background was found in ref. \cite{LuRoy} by means of an
$S$-duality transformation of a solution in which only one of the two
three-form field strengths is non-vanishing. By adapting the results of
ref. \cite{LuRoy}, it is straightforward to conclude that the matrix
$\Lambda$ suited for our objectives has the following elements:
\bear
a&=&-e^{\phi_0}\,(q\,+\,p\chi_0\,)\,
\Bigl[\,\rho_{(p,q)}\,\Bigr]^{-{1\over 2}}\,\,,\rc
b&=&\Bigl(\,
e^{\phi_0}\,(q\,+\,p\chi_0\,)\,\chi_0\,+\,e^{-\phi_0}\,p\,\Bigr)\,
\Bigl[\,\rho_{(p,q)}\,\Bigr]^{-{1\over 2}}\,\,,\rc
c&=&-p\,\Bigl[\,\rho_{(p,q)}\,\Bigr]^{-{1\over 2}}\,\,,\rc
d&=&-q\,\Bigl[\,\rho_{(p,q)}\,\Bigr]^{-{1\over 2}}\,\,.\rc
\label{tnueve}
\eear
Let us call ${\cal M}^{(D)}$ to the transformed ${\cal M}$ matrix, \ie:
\beq
{\cal M}^{(D)}\,=\,\Lambda\,{\cal M}\,\Lambda^T\,\,.
\label{cuarenta}
\eeq
After some algebra, one can check that, when the matrix elements of
$\Lambda$ are those displayed in eq. (\ref{tnueve}), the dual matrix 
$\cal M^{(D)}$ has the simple form:
\beq
\cal M^{(D)}\,=\,
\pmatrix{\Bigl[\,H_{(p,q)}(r)\,\Bigr]^{{1\over 2}}&0\cr
         0&\Bigl[\,H_{(p,q)}(r)\,\Bigr]^{-{1\over 2}}}\,\,.
\label{cuno}
\eeq
This matrix  corresponds to the following values of the dilaton and RR
scalar:
\beq
e^{-\phi^{(D)}}\,=\,\Bigl[\,H_{(p,q)}(r)\,\Bigr]^{{1\over 2}}\,\,,
\,\,\,\,\,\,\,\,\,\,\,\,\,\,\,\,\,\,\,\,\,\,\,\,\,\,
\chi^{(D)}\,=\,0\,\,,
\label{cdos}
\eeq
which are the values of $\phi$ and $\chi$ for a D5-brane solution of
type IIB supergravity. Moreover, after the transformation, the vector
${\cal D}$ takes the form:
\beq
{\cal D}^{(D)}\,=\,
\big(\,\Lambda\,^{-1})^T\,{\cal D}\,=\,
\pmatrix{-\,{\cal F}_{12}^{(D)}\cr\cr
         \Pi^{(D)}\cr}\,\,,
\label{ctres}
\eeq
where:
\beq
{\cal F}_{12}^{(D)}\,=\,d\,{\cal F}_{12}\,+\,c\,\Pi\,\,,
\,\,\,\,\,\,\,\,\,\,\,\,\,\,\,\,\,\,\,\,\,\,\,\,\,\,
\Pi^{(D)}\,=\,a\,\Pi\,+\,b\,{\cal F}_{12}\,\,.
\label{ccuatro}
\eeq

In terms of the dual variables, the energy functional is simply given
by:
\beq
U\,=\,T_3\,\int\,d^3\theta\,
\sqrt{r^2\,+\,r^{\,'\,2}}\,\,
\sqrt{\Sigma_{(p,q)}^2\,+\,\Big(\,\Pi^{(D)}\,\Big)^2
\,+\,H_{(p,q)}(r)\,\Big(\,{\cal F}_{12}^{(D)}\,\Big)^2}\,\,.
\label{ccinco}
\eeq

From eq. (\ref{ccuatro}) and the Gauss law and Bianchi identity for
the original variables $\Pi$ and ${\cal F}_{12}$ (eqs. (\ref{vcuatro})
and (\ref{vseis})), it is immediate to obtain the equation
that determines the dependence on $\theta$ of ${\cal F}_{12}^{(D)}$,
namely:
\beq
\partial_{\theta}\,{\cal F}_{12}^{(D)}\,=\,
(\,qc\,-\,pd\,)\,2\,N\,\alpha'\,\,\sqrt{\bar g}\,\,.
\label{cseis}
\eeq
By using the explicit values of the coefficients $d$ and $c$ from eq. 
(\ref{tnueve}), one can verify that the right-hand side of eq.
(\ref{cseis}) vanishes, \ie:
\beq
\partial_{\theta}\,{\cal F}_{12}^{(D)}\,=\,0\,\,.
\label{csiete}
\eeq
Eq. (\ref{csiete}) is simply the Bianchi identity for the dual
variables. As the dual background has vanishing NSNS gauge field, the
result of eq. (\ref{csiete}) was to be expected. Moreover, it follows
from the analysis of ref. \cite{baryon} that the D3-brane
configurations which minimize the energy in a D5 background are such
that the magnetic components of the world-volume field strength are
zero. Thus, in order to solve the equations of motion derived from $U$, 
we shall adopt the ansatz:
\beq
{\cal F}_{12}^{(D)}\,=\,0\,\,.
\label{cocho}
\eeq
If this condition holds,  it implies from (\ref{ccuatro}) the following
relation between  ${\cal F}_{12}$ and $\Pi$:
\beq
{\cal F}_{12}\,=\,-{c\over d}\,\Pi\,=\,-{p\over q}\,\Pi\,\,,
\label{cnueve}
\eeq
which is, in principle, valid only when $q\not= 0$ (we will consider
the $q=0$ case below). Using the relation (\ref{cnueve}) in eq. 
(\ref{ccuatro}), we can get $\Pi^{(D)}$ in terms of $\Pi$, namely:
\beq
\Pi^{(D)}\,=\,-\,
{\Bigl[\,\rho_{(p,q)}\,\Bigr]^{{1\over 2}}\over q}\,\,\Pi\,\,.
\label{cincuenta}
\eeq
Let us now determine $\Pi$ by integrating the Gauss law constraint 
(\ref{vcuatro}). Taking eq. (\ref{dsiete}) into account, one can solve
eq. (\ref{vcuatro}) as follows:
\beq
\Pi\,=\,2q\,N\,\alpha'\,\sin\theta^2\,\,b_{\nu}(\theta)\,\,,
\label{ciuno}
\eeq
where $b_{\nu}(\theta)$ satisfies the differential equation 
(\ref{veinte}). It is immediate to verify that:
\beq
b_{\nu}(\theta)\,=\,{1\over 2}\,
[\,\theta\,-\,\cos\theta\sin\theta\,-\,\pi\nu\,]\,\,,
\label{cidos}
\eeq
is the general solution of eq. (\ref{veinte}). In eq. (\ref{cidos})
$\nu$ is an integration constant which, for example, determines the
value of $b_{\nu}(\theta)$ at $\theta=\pi$. By plugging in eq.
(\ref{cincuenta}) the expression of  $\Pi$  given by (\ref{ciuno}), one
can obtain $\Pi^{(D)}$ as a function of the angles. One gets:
\beq
\Pi^{(D)}\,=\,
\,=\,-2\,R_{(p,q)}^2\,\sin\theta^2\,b_{\nu}(\theta)\,\,.
\label{citres}
\eeq
Similarly, from eqs. (\ref{cnueve}) and (\ref{ciuno}) we can determine 
${\cal F}_{12}$:
\beq
{\cal F}_{12}\,=\,-2pN\alpha'\,\sin\theta^2\,b_{\nu}(\theta)\,\,.
\label{cicuatro}
\eeq
It is immediate to verify that, indeed, the expression of 
${\cal F}_{12}$ given in eq. (\ref{cicuatro}) solves the Bianchi
identity (\ref{vseis}). Actually, we could have obtained 
${\cal F}_{12}$ by direct integration of eq.~(\ref{vseis}). The
advantage of our present derivation based on $S$-duality is that it
fixes uniquely the integration constant of this equation. Moreover, if
one chooses the gauge for $B$ as in eq. (\ref{dnueve}), \ie\ with the
same constant $\nu$ as is in eq. (\ref{cicuatro}), one has:
\beq
{\cal F}_{12}\,=\,- B_{\theta^1\theta^2}\,\,,
\label{cicinco}
\eeq
which implies that:
\beq
F_{\theta^1\theta^2}\,=\,0\,\,.
\label{ciseis}
\eeq
Notice that eq. (\ref{ciseis}) implies that  the
spatial components $A_{\theta^i}$ of the world-volume gauge field
vanish and, therefore, only $A_0$ is non-zero and depends on the polar
angle $\theta$.

Let us now define the function $D_{(p,q)}(\theta)$ as:
\beq
D_{(p,q)}(\theta)\,\equiv\,-2\,R_{(p,q)}^2\,b_{\nu}(\theta)\,\,.
\label{cisiete}
\eeq

The field strength ${\cal F}_{12}$ and the dual momentum $\Pi^{(D)}$
 can be rewritten  in terms of $D_{(p,q)}(\theta)$. Indeed, by using
eq. (\ref{seis}), one arrives at:
\beq
{\cal F}_{12}\,=\,{p\over [\rho_{(p,q)}]^{{1\over 2}}}\,\,
\sin\theta^2\,\,D_{(p,q)}(\theta)\,\,,
\label{ciocho}
\eeq
while  $\Pi^{(D)}$ is given by:
\beq
\Pi^{(D)}\,=\,\sin\theta^2\,D_{(p,q)}(\theta)\,\,. 
\label{cinueve}
\eeq

Let us now consider the problem of finding the embeddings of the
D3-brane with minimal energy $U$. First of all, we can substitute 
in the expression of $U$ written in eq. (\ref{ccinco}) 
the values of ${\cal F}_{12}^{(D)}$ and $\Pi^{(D)}$ that we have
obtained  (eqs. (\ref{cocho}) and (\ref{cinueve})). As, by
assumption, $r$ only depends on the polar angle $\theta$, it is
possible to integrate over the other two  angles $\theta^1$ and
$\theta^2$, with the result:
\beq
U\,=\,4\pi\,T_3\,\,\int d\theta\,
\sqrt{r^2\, +\,r\,'\,^{2}}\,\,
\sqrt{[\,\Delta_{(p,q)}(r,\theta)\,]^2\,+\,[\,D_{(p,q)}(\theta)\,]^2}
\,\,,
\label{sesenta}
\eeq
where:
\beq
\Delta_{(p,q)}(r,\theta)\,\equiv\,r^2\,H_{(p,q)}(r)\,
(\,\sin\theta\,)^2\,\,.
\label{ssuno}
\eeq
Eq. (\ref{sesenta}) will be the starting point in our determination of
the minimal energy configurations. Before proceeding further, let us
point out that eq. (\ref{sesenta})  is also valid when $q=0$, \ie\
when the background is purely NS. Actually, although some of our
intermediate equations are ill-defined when $q=0$, the final
expressions for $\Pi$ and ${\cal F}_{12}$ (eqs. (\ref{ciuno})  and 
(\ref{cicuatro}) respectively) are perfectly well-behaved in this
case,  and one can readily check that they solve the Gauss law and the
Bianchi identity for $q=0$. By substituting these $q=0$ values of 
$\Pi$ and ${\cal F}_{12}$ in eq. (\ref{treinta}),  one can easily
conclude that the energy $U$ is given by eq. (\ref{sesenta}) also when 
$q=0$. Therefore,  we will assume that this value is included
in what follows. 

The minimal energy embeddings of the D3-brane are characterized by a
function $r(\theta)$ which satisfies the Euler-Lagrange equation of
motion derived from $U$. This equation of motion is a rather
complicated second-order differential equation which, in general, we
will not be able  to solve. Instead, we will find a first-order
differential equation for $r(\theta)$ whose solutions saturate an
energy bound and, thus, also solve the Euler-Lagrange equation. In
order to find this first-order equation we follow closely refs.
\cite{Craps, baryon}. First of all, we define the function
$f_{(p,q)}(r,\theta)$ as follows:
\beq
f_{(p,q)}(r,\theta)\,\equiv\,
{\Delta_{(p,q)}(r,\theta)\sin\theta\,+\,D_{(p,q)}(\theta)\cos\theta
\over 
\Delta_{(p,q)}(r,\theta)\cos\theta\,-\,D_{(p,q)}(\theta)\sin\theta}
\,\,.
\label{suno}
\eeq
In terms of $f_{(p,q)}(r,\theta)$ it is easy to check that the square
root in $U$ can be written as a sum of squares. One has:
\bear
U\,&=&\,
4\pi\,T_{3}\,\,\int d\theta\,
\,\,\times\rc\rc
&&\times\,\sqrt{{\cal Z}^2\,+\,r^2\,
\Bigl[\,\Delta_{(p,q)}(\,r\,,\,\theta\,)\,
{\rm cos }\,\theta-\,
D_{(p,q)}(\theta)\,{\rm sin }\,\theta\,\Bigr]^2\,\Bigl[\,
{r'\over r}-f_{(p,q)}(r, \theta)\,\Bigr]^2
}\,\,,
\label{sdos}
\eear
where ${\cal Z}$ is given by:
\beq
{\cal Z}\,=\,r\,
\Bigl[\,\Delta_{(p,q)}(\,r\,,\,\theta\,)\,
{\rm cos }\,\theta-\,
D_{(p,q)}(\theta)\,{\rm sin }\,\theta\,\Bigr]\,
\Bigl[\,1\,+\,{r'\over r}\,f_{(p,q)}(r, \theta)\,\Bigr]\,\,.
\label{stres}
\eeq
From eq. (\ref{sdos}) we obtain immediately that $U$ satisfies the
bound:
\beq
U\,\ge\,4\pi\,T_{3}\,\int d\theta\,
\Bigl|\,{\cal Z}\,\Bigr|\,\,,
\label{scuatro}
\eeq
which is saturated when $r(\theta)$ satisfies the following
first-order differential equation:
\beq
{r'\over r}\,=\, f_{(p,q)}(r,\theta)\,\,.
\label{scinco}
\eeq
Eq. (\ref{scinco}) is the first-order differential equation we were
looking for. It is easy to prove that any function $r(\theta)$ that
satisfies (\ref{scinco}) also solves the Euler-Lagrange equation.
Moreover, ${\cal Z}$ can be written as a total derivative:
\beq
{\cal Z}\,=\,
{d\over d\theta}\,\,
\Bigl[\,D_{(p,q)}(\theta)\,r\,{\rm cos }\,\theta\,+\,
\Bigl(\,{1\over 3}\,+\,{R^{2}_{(p,q)}\over r^{2}}\,\Bigr)
\,\,(\,r\,{\rm sin }\,\theta\,)^{3}\,\Bigr]\,\,.
\label{sseis}
\eeq
Eq. (\ref{sseis}) holds for  an arbitrary  function $r(\theta)$. In
fact, in order to demonstrate it,  one only has to use the value of the
derivative of $D_{(p,q)}(\theta)$, which can be easily obtained from
its explicit expression (eq. (\ref{cisiete})). The main consequence of
eq. (\ref{sseis}) is that only the boundary values of $r(\theta)$
contribute to the energy bound of eq. (\ref{scuatro}) and, thus, the
solutions of (\ref{scinco}) certainly minimize the energy for given
boundary conditions. We will refer to eq. (\ref{scinco}) as the BPS
condition. In section 5 we will reobtain this condition as the one
needed for preservation of 1/4 of supersymmetry. In order to establish
this connection with supersymmetry, it is interesting to recast the
BPS condition as an ansatz for the electric components $F_{0\theta}$
of the world-volume field strength. Recall that $F_{0\theta}$, as a
function of $\Pi$ and ${\cal F}_{12}$ was given in eq.
(\ref{vsiete}). Actually, by using the values of the dilaton and RR
scalars in the $(p,q)$ five-brane background (eqs. (\ref{nueve}) and
(\ref{diez})), together with the expressions of $\Pi$ and 
${\cal F}_{12}$ (eqs. (\ref{ciuno}) and
(\ref{cicuatro})), one can prove, after some calculation, that:
\beq
e^{\phi}\,\Big(\,\Pi\,-\,\chi\,{\cal F}_{12}\,)\,=\,
-\Big[\,H_{(p,q)}(r)\,\Big]^{-{1\over2}}\,\,\,
{q+p\chi_0\over \Big[\,\rho_{(p,q)}\,\Big]^{{1\over 2}}}\,\,\,
e^{\phi_0}\,\,\sin\theta^2\,\,\,D_{(p,q)}(\theta)\,\,.
\label{ssiete}
\eeq
By using eq. (\ref{ssiete}) to determine the right-hand side of eq. 
(\ref{vsiete}), it is not difficult to find the following expression
for $F_{0\theta}$:
\beq
F_{0\theta}\,=\,-
{q+p\chi_0\over \Big[\,\rho_{(p,q)}\,\Big]^{{1\over 2}}}\,\,
e^{\phi_0}\,\,
{\sqrt{r^2\,+\,r'^{\,2}}\over 
\sqrt{\,\Big[\,\Delta_{(p,q)}(r,\theta)\,\Big]^2\,+\,
\Big[\,D_{(p,q)}(\theta)\,\Big]^2\,}}\,\,D_{(p,q)}(\theta)\,\,.
\label{socho}
\eeq
Moreover, if $r(\theta)$ satisfies the BPS condition (\ref{scinco}),
one can readily check that:
\beq
{\sqrt{r^2\,+\,r'^{\,2}}\over 
\sqrt{\,\Big[\,\Delta_{(p,q)}(r,\theta)\,\Big]^2\,+\,
\Big[\,D_{(p,q)}(\theta)\,\Big]^2\,}}\,\,D_{(p,q)}(\theta)\,\,=\,
(\,r\cos\theta\,)'\,\,,
\label{snueve}
\eeq
which implies that, when the BPS condition is fulfilled, the electric
field $F_{0\theta}$ takes the value:
\beq
F_{0\theta}\,=\,
-{q+p\chi_0\over \Big[\,\rho_{(p,q)}\,\Big]^{{1\over 2}}}\,\,
e^{\phi_0}\,\,
(\,r\cos\theta\,)'\,\,.
\label{setenta}
\eeq
Notice that, as  $F_{0\theta}\,=\,-\partial_{\theta}\,A_0$, the time
component of the world-volume gauge field is related to $r(\theta)$ as:
\beq
A_{0}(\theta)\,=\,
{q+p\chi_0\over \Big[\,\rho_{(p,q)}\,\Big]^{{1\over 2}}}\,\,
e^{\phi_0}\,\,
r(\theta)\cos\theta\,\,.
\label{stuno}
\eeq
It is also interesting to relate ${\cal F}_{12}$  to $r(\theta)$ and
its derivative. This can be done by rewriting the BPS condition 
(\ref{scinco}) as:
\beq
D_{(p,q)}(\theta)\,=\,
{(r\cos\theta)'\over (r\sin\theta)'}\,\,\Delta_{(p,q)}(r,\theta)\,\,.
\label{stdos}
\eeq
After substituting $D_{(p,q)}(\theta)$, as given in eq. (\ref{stdos}), 
on the right-hand side of eq.  (\ref{ciocho}), one gets:
\beq
{\cal F}_{12}\,=\,
{p\over \Big[\,\rho_{(p,q)}\,\Big]^{{1\over 2}}}\,\,
\sin\theta^2\,\,\,
{(r\cos\theta)'\over (r\sin\theta)'}\,\,\Delta_{(p,q)}(r,\theta)\,\,.
\label{sttres}
\eeq
We shall demonstrate in section 5 that, in order to preserve 1/4
supersymmetry, $F_{0\theta}$ and ${\cal F}_{12}$ must be related to
$r(\theta)$ as in eqs. (\ref{setenta}) and (\ref{sttres}). This result
will clarify the nature of our first-order equation (\ref{scinco}) .

An amazing aspect of eq. (\ref{scinco}) is the fact that, as was
noticed in ref. \cite{baryon}, it can be integrated exactly. In fact,
by using the expression of $D_{(p,q)}(\theta)$ (eq. (\ref{cisiete})),
one can  rewrite eq. (\ref{scinco})  as a total derivative:
\beq
{d\over d\theta}\,(r\cos\theta)\,=\,R_{(p,q)}^2\,
{d\over d\theta}\,\Big[\,{\theta-\pi\nu\over r\sin\theta}
\,\Big]\,\,.
\label{stcuatro}
\eeq
The integration of eq. (\ref{stcuatro}) is immediate and the result
can be written as:
\beq
r\cos\theta\,=\,R_{(p,q)}^2\,\,{\theta-\pi\nu\over r\sin\theta}
\,-\,z_{\infty}\,\,,
\label{stcinco}
\eeq
where $z_{\infty}$ is a constant of integration. Notice that 
(\ref{stcinco}) determines $r(\theta)$ in implicit form. However, from
eq. (\ref{stcinco}) one can easily obtain the explicit expression of
$r(\theta)$. The result is:
\beq
r(\theta)\,=\,-{z_{\infty}\over 2\cos\theta}\,\pm\,
{1\over 2\cos\theta}\,
\sqrt{z_{\infty}^2\,+\,4\,R_{(p,q)}^2\,\,
{\theta-\pi\nu\over \tan\theta}}\,\,.
\label{stseis}
\eeq
In figure 1 we have represented the D3-brane embedding for different
values of the constants $\nu$ and $z_{\infty}$. In order to choose
the signs on the right-hand side of (\ref{stseis}) one has to  take into
account that $r(\theta)$ must be real and non-negative.
Notice that, in general, the function $r(\theta)$ is double-valued due
to these two signs.  This
double-valueness can also be appreciated from the plots of figure 1.
In order to have a global description of the D3-brane world-volume by
means of a single-valued function, it is more convenient to work in a
new set of variables $(z,\rho)$, related to $(r,\theta)$ as follows:
\beq
z\,=\,-r\,{\rm cos }\,\theta\,\,,
\,\,\,\,\,\,\,\,\,\,\,\,\,\,\,\,\,\,\,\,\,\,\,\,\,\,\,\,
\rho\,=\,r\,{\rm sin }\,\theta\,\,.
\label{stsiete}
\eeq
Notice that $z$ and $\rho$ are nothing but cylindrical coordinates.
Clearly,  $z$ can take values on the interval $(-\infty, +\infty)$,
whereas $0\le\rho<\infty$. In these coordinates the embedding of the
D3-brane is determined by a single-valued function $z(\rho)$. 
\begin{figure}
\centerline{\epsffile{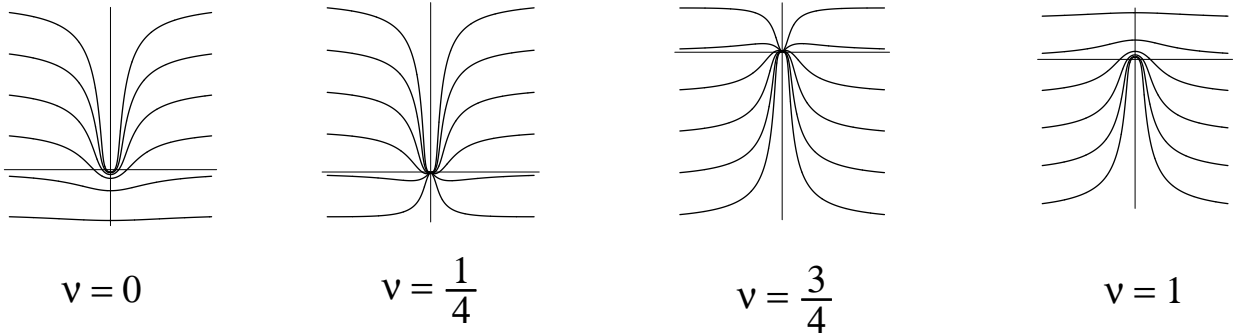}}
\caption{Solutions of the BPS differential equation for
several values of $\nu$ and $z_{\infty}$. The coordinates $(r,\theta)$
are the polar  coordinates of the plane of the figure ($\theta =\pi$
at the top). The horizontal and vertical axis correspond,
respectively, to the coordinates $\rho$ and $z$. For every value of
$\nu$, the different curves plotted together correspond to different
values of the constant $z_{\infty}$.}
\label{fig1}
\end{figure}

The solution (\ref{stcinco}) of the BPS differential
equation was analyzed in detail in ref. \cite{baryon}. Let us review
here the main results obtained in \cite{baryon}. First of all, in
those regions which are far from the origin $r=0$ (or
$\rho\rightarrow\infty$ in cylindrical coordinates) the D3-brane
world-volume can be described approximately by the equation $z={\rm
constant}=z_{\infty}$. Thus, in these asymptotic regions, the shape of
the D3-brane world-volume is just a plane, \ie\ the brane probe is not
bent by the action of the background. For some values of the
integration constants the shape of the D3-brane near the polar axis
(\ie\ for $\theta=0$ or $\theta=\pi$) resembles that of a tube (see
figure 1). In this region, the D3-brane develops a world-volume
soliton which is a spike that connects the D3-brane probe to the $r=0$
region, where the $(p,q)$ five-branes of the background are located. In
order to find out what this soliton represents, let us evaluate its
energy. For this purpose it is more convenient to express the energy
functional $U$ in cylindrical coordinates. By performing the change of
variables (\ref{stsiete}) on the right-hand side of eq.
(\ref{sesenta}), one gets the remarkably simple result \cite{baryon}:
\beq
U\,=\,4\pi\,T_{3}\,\,\int d\rho\,\,
\sqrt{\Bigl(\,\,\Delta_{(p,q)}\,
-\,D_{(p,q)}\,{dz\over d\rho}\,\,\Bigr)^2\,+\,
\Bigl(\,\,D_{(p,q)}
\,+\,\Delta_{(p,q)}\,{dz\over d\rho}\,\,\Bigr)^2}\,\,.
\label{stocho}
\eeq
Moreover, the BPS condition in the $(z,\rho)$ variables takes the form:
\beq
{dz\over d\rho}\,=\,-{D_{(p,q)}\over \Delta_{(p,q)}}\,\,.
\label{stnueve}
\eeq
Notice that, when $z(\rho)$ satisfies eq. (\ref{stnueve}), the second
term under the square root on the right-hand side of eq.
(\ref{stocho}) vanishes. Therefore, the energy of a BPS configuration
is simply:
\beq
U_{BPS}\,=\,-\,4\pi\,T_{3}\,\,\int d\rho\,\,
\Bigl[\,{dz\over d\rho}\,+\, 
\Bigl(\,{dz\over d\rho}\,\Bigr)^{-1}\,\Bigr]\,\,D_{(p,q)}\,\,\,\,.
\label{ochenta}
\eeq
Notice that the argument of the function $D_{(p,q)}$ in
(\ref{ochenta}) is $\theta=\,-\,\arctan(-\rho/z)$. From
eq.~(\ref{ochenta}) it is very easy to evaluate the contribution of the
soliton to the energy $U$. Indeed, as the shape of the soliton is
approximately a cylinder with $\rho={\rm constant}$, the derivative 
${dz\over d\rho}$ is very large and, thus, one can neglect in
(\ref{ochenta})  the term containing 
$\Bigl(\,{dz\over d\rho}\,\Bigr)^{-1}$. Moreover, the argument of the
function  $D_{(p,q)}$  is almost constant and equal to $0$ or $\pi$
and, therefore, $D_{(p,q)}$  can be taken out of the integral. Thus,
for a tube at $\theta=\pi$, the energy is given by:
\beq
U_{tube}\,=\,4\pi\,T_3\,
|\,D_{(p,q)}(\pi)\,|\,L_{tube}\,\,,
\label{ouno}
\eeq
where $L_{tube}$ is the length of the tube defined as:
\beq
L_{tube}\,=\,\int_{tube}\,d\rho\,\Bigr|{dz\over d\rho}\Bigl|\,\,.
\label{odos}
\eeq

It is clear that the world-volume soliton we have found is
one-dimensional, \ie\ is a string-like object. The best way to identify
it is by computing its tension and comparing the result with the
strings of the type IIB theory. This tension is naturally defined in
the string frame, where the metric $g_{string}$ is related to the
Einstein metric $g$ we have been using as 
$g_{string}\,=\,e^{\phi/2}\,g$. The world-volume action of any string
has a Nambu-Goto term which contains the square root of the determinant
of the induced metric. This  Nambu-Goto term is multiplied by a factor
$e^{\phi/2}$ when one changes from the string frame to the Einstein
frame. It is, thus,  clear that this same factor, evaluated at infinity,
appears in the relation of the string tensions in the two frames.
Actually, the tension in the string frame is obtained by multiplying by 
$g_s^{-1/2}$ the tension evaluated in the Einstein frame. Taking these
considerations into account, we obtain from eq. (\ref{ouno}) that the
tension of our soliton in the string frame is:
\beq
g_s^{-{1\over 2}}\,4\pi\,T_3\,
|\,D_{(p,q)}(\pi)\,|\,\,.
\label{otres}
\eeq
The value of $D_{(p,q)}$ at $\theta=\pi$ can be obtained in an
straightforward way from eqs. (\ref{cisiete}) and (\ref{cidos}):
\beq
|\,D_{(p,q)}(\pi)\,|\,=\,\pi\,(\,1\,-\,\nu\,)\,R_{(p,q)}^2\,\,,
\label{ocuatro}
\eeq
where, for reasons which will become clear in a moment, we have taken
$\nu$ in the interval $0\le\nu\le 1$. By using the value of
$R_{(p,q)}$, as given in eq. (\ref{seis}), one gets that the tension
of the spike is:
\beq
(1-\nu)\,N\,T_{f}\,
\Bigl[\,{\rho_{(p,q)}\,\over g_s}\Bigr]^{{1\over 2}}\,\,.
\eeq
Recalling \cite{Schwarz} that the tension of a $(p,q)$ string is given
by:
\beq
T_{(p,q)}\,=\,\sqrt{\,(p-q\chi_0)^2\,+\,{q^2\over g_s^2}}\,\,\,
T_{f}\,=\,\Bigl[\,{\rho_{(q,-p)}\,\over g_s}\Bigr]^{{1\over 2}}
\,\,T_{f}\,\,,
\label{oseis}
\eeq
we conclude that  the tension of our  spike is:
\beq
(1\,-\,\nu)\,N\,T_{(-q,p)}\,\,,
\label{osiete}
\eeq
which corresponds to a bundle constituted by  $(1\,-\,\nu)\,N$
parallel $(-q,p)$-strings. It is now clear that $\nu$ should be
quantized in units of $1/N$ and restricted to take values in the
interval $[0,1]$. 

The analysis we have done can also be carried out  for
world-volume solitons at $\theta=0$. The conclusion, in this case, is
that the spike can be interpreted as a bundle of $\nu\,N$
parallel $(-q,p)$-strings. Therefore, the configuration we have found
involves three branes, background, probe and soliton,
intersecting according to the array:
\beq
\ba{ccccccccccl}
(p,q)\,\,\, {\rm five-brane}: &1&2&3&4&5&\_&\_&\_&\_ &\quad
\mbox{background}   
\nonumber\\ 
D3: &\_&\_&\_&\_&\_&6&7&8&\_ &\quad \mbox{probe}     \nonumber \\
(-q,p)\,\,\, {\rm  string}: &\_&\_&\_&\_&\_&\_&\_&\_&9 &\quad
\mbox{soliton.}
\ea
\label{oocho}
\eeq
The $9$-direction is the radial one.

The world-volume solitons we have found provide an explicit
realization of the Hanany-Witten effect \cite{HW}. In order to
clarify this point, let us consider, for concreteness, the $\nu=0$
case. As can be seen from figure 1, when
$z_{\infty}\rightarrow-\infty$, the brane is nearly flat. When
$z_{\infty}$ is increased, the D3-brane gets more and more deformed
near the $\theta\,=\,\pi$ axis,  and,  when
$z_{\infty}\rightarrow+\infty$ the bundle of $(-q,p)$-strings is
created. For $\nu=1$ the soliton is created when
$z_{\infty}\rightarrow-\infty$ and the brane is not bent when 
$z_{\infty}\rightarrow+\infty$. Notice that for $\nu=0,1$, the
D3-brane does not reach $r=0$ and only in those two cases the D3-brane
wraps completely the sphere $S^3$. If, on the contrary, $\nu\not=0,1$,
the $r=0$ point is reached and the solution has  the two types of
soliton tubes (at $\theta=0,\pi$), depending on the sign of
$z_{\infty}$.

\bigskip
\setcounter{equation}{0}
\section{Supersymmetry}
\medskip
In this section we will show that the world-volume configurations we
have found preserve $1/4$ of the background supersymmetry. As is
well-known \cite{swedes, bbs}, the number of supersymmetries preserved
by a D3-brane moving in the background of a $(p,q)$ five-brane is the
number of independent solutions of the equation:
\beq
\Gamma_{\kappa}\,\epsilon_{(p,q)}\,=\,\epsilon_{(p,q)}\,\,,
\label{onueve}
\eeq
where $\Gamma_{\kappa}$ is the matrix appearing in the
$\kappa$-symmetry transformation of the D3-brane and 
$\epsilon_{(p,q)}$ is a  Killing spinor of the background. In general,
the matrix $\Gamma_{\kappa}$ depends on the background and on the type
of brane. For a D3-brane in a type IIB background,  
$\Gamma_{\kappa}$ is given by:
\bear
\Gamma_{\kappa}\,&=&\,{1\over 
\sqrt{-{\rm det}\,(\,g\,+\,e^{-{\phi\over 2}}\,{\cal F}\,)}}\,\,
\sum_{n=0}^{\infty}\,{1\over 2^n\,n!}\,\,
\gamma^{\mu_1\nu_1\cdots\mu_n\nu_n}\times\rc\rc\rc
&&\times\,e^{-{\phi\over 2}}\,{\cal F}_{\mu_1\nu_1}\,\cdots
e^{-{\phi\over 2}}{\cal F}_{\mu_n\nu_n}\,\,J^{(n)}\,\,,
\label{noventa}
\eear
where $J^{(n)}$ is the following matrix:
\beq
J^{(n)}\,=\,(-1)^n\,\sigma_3^n\,(i\sigma_2)\,\otimes\,\Sigma_0\,\,,
\label{nuno}
\eeq
with $\Sigma_0$ being:
\beq
\Sigma_0\,=\,{1\over 4!}\,\,\epsilon^{\mu_1\cdots\mu_4}\,
\gamma_{\mu_1\cdots\mu_4}\,\,.
\label{ndos}
\eeq
Notice that the presence of the $e^{-{\phi\over 2}}$ factors on the
right-hand side of eq. (\ref{noventa}) is due to the fact that $g$ is
the induced metric in the Einstein frame. Moreover, in eqs.
(\ref{noventa}) and (\ref{ndos}) $\gamma_{\mu\nu\cdots}$ are
antisymmetrized products of the induced world-volume Dirac matrices
which, in terms of the ten-dimensional constant gamma matrices 
$\Gamma_{\underline{m}}$ are given by:
\beq
\gamma_{\mu}\,=\,
\partial_{\mu}\,X^m\,E_{m}^{\underline{m}}\,
\Gamma_{\underline{m}}\,\,,
\label{ntres}
\eeq
where $E_{m}^{\underline{m}}$ is the ten-dimensional vielbein.

The infinite sum appearing on the right-hand side of eq.
(\ref{noventa}) has, actually, only three terms due to the
antisymmetry of the index structure of this equation, and, thus,
$\Gamma_{\kappa}$  can be written as:
\bear
\Gamma_{\kappa}\,=\,{1\over 
\sqrt{-{\rm det}\,(\,g\,+\,e^{-{\phi\over 2}}\,{\cal F}\,)}}\,\,
(i\sigma_2)\,&\Bigg[&\,\Sigma_0\,-\,{1\over 4}\,\sigma_3\,
\epsilon^{\mu\nu\rho\sigma}\,e^{-{\phi\over 2}}\,
{\cal F}_{\mu\nu}\,\gamma_{\rho\sigma}\,+\,\rc\rc\rc
&&+\,{1\over 8}\,\epsilon^{\mu\nu\rho\sigma}\,e^{-{\phi\over 2}}\,
{\cal F}_{\mu\nu}\,\,e^{-{\phi\over 2}}\,
{\cal F}_{\rho\sigma}\,\,\Bigg]\,\,.
\label{ncuatro}
\eear
In eq. (\ref{ncuatro}), and in what follows,  we have suppressed the
tensor product symbol. For a metric of the form  (\ref{cuatro}), and if
only the radial coordinate $r$ is excited, the induced gamma matrices 
take the form:
\bear
\gamma_0\,&=&\,[\, H_{(p,q)}(r)\,]^{-{1\over 8}}\,
\,\Gamma_{\underline 0}\,\,,\rc\rc
\gamma_{\theta^i}\,&=&\,[\, H_{(p,q)}(r)\,]^{{3\over 8}}\,\,
\Big[\,r\,e_i^{\underline i}\,\Gamma_{\underline{\theta^i}}\,+\,
\partial_{\theta^i}\,r\,\Gamma_{\underline r}\,\Big]\,\,,
\label{ncinco}
\eear
where $e_i^{\underline i}$ are the vielbeins on the sphere $S^3$.
Moreover,  when $r$ only depends on $\theta^3\equiv\theta$, the matrix
$\Sigma_0$ can be written as:
\beq
\Sigma_0\,=\,r^2\, H_{(p,q)}(r)\,\sqrt{\bar g}\,
[\,r'\,\Gamma_{\underline{\theta^3}}\,-\,r\Gamma_{\underline{r}}\,]\,
\Gamma_{\underline{0r\theta^1\theta^2\theta^3}}\,\,.
\label{nseis}
\eeq
In agreement with our analysis of section 4, we shall assume from now
on that the only non-vanishing components of ${\cal F}_{\mu\nu}$ are 
${\cal F}_{0\theta}$ and ${\cal F}_{\theta^1\theta^2}$.
If we define:
\bear
f_{0\theta}\,&\equiv&\,{e^{-{\phi\over 2}}\,{\cal F}_{0\theta}\over
[\, H_{(p,q)}(r)\,]^{{1\over 4}}}\,\,,\rc\rc
f_{\theta^1\theta^2}\,&\equiv&\,{e^{-{\phi\over 2}}
\,{\cal F}_{\theta^1\theta^2}
\over r^2\,[\, H_{(p,q)}(r)\,]^{{3\over 4}}\sqrt{\bar g}}\,\,,
\label{nsiete}
\eear
then, after using eqs. (\ref{ncuatro})-(\ref{nseis}), we can write eq.
(\ref{onueve}) as:
\bear
&&\sqrt{(\,r^2\,+\,r'^2\,-\, f_{0\theta}^2\,)\,
(1\,+\,f_{\theta^1\theta^2}^2\,)}\,\,\,\epsilon_{(p,q)}\,=\,\rc\rc
&&=i\sigma_2\,\Big[\,\sigma_3\,(\,r'\,\Gamma_{\underline{0r}}\,+\,
r\Gamma_{\underline{0\theta}}\,)\,-\,f_{0\theta}\,\,\Big]\,\,
\Big[\,\sigma_3\,\Gamma_{\underline{\theta^1\theta^2}}\,-\,
f_{\theta^1\theta^2}\,\Big]\,\,\,\epsilon_{(p,q)}\,\,.
\label{nocho}
\eear

We shall regard (\ref{nocho}) as an equation whose unknowns are both
the spinors $\epsilon_{(p,q)}$ and the world-volume field strength 
${\cal F}$. In order to solve this equation we must know first the
general form of the Killing spinors of the $(p,q)$ five-brane 
background. These spinors can be obtained from the supersymmetric
variation of the gravitino and dilatino fields in the type IIB
supergravity configurations of eqs. (\ref{cuatro})-(\ref{once}). In
the coordinate system in which we are working, and in the Einstein
frame, the  $\epsilon_{(p,q)}$'s can be written as
follows \cite{Lls}:
\beq
\epsilon_{(p,q)}\,=\,
{1\over[\, H_{(p,q)}(r)\,]^{{1\over 16}}}\,\,
e^{-{i\over 2}\alpha_{(p,q)}(r)\,\sigma_2}\,\,
e^{-{\theta^3\over 2}\,\Gamma_{\underline{\theta^3 r}}}\,
e^{-{\theta^2\over 2}\,\Gamma_{\underline{\theta^2 \theta^3}}}\,
e^{-{\theta^1\over 2}\,\Gamma_{\underline{\theta^1 \theta^2}}}\,
\epsilon_{0}\,\,,
\label{nnueve}
\eeq
where $\epsilon_0$ is a constant spinor of positive ten-dimensional
chirality ($\Gamma_{11}\,\epsilon_{0}\,=\,\epsilon_{0}$), satisfying
the condition:
\beq
\Gamma_{\underline 0}\,\Gamma_{\underline {12345}}\,
\sigma_1\,\epsilon_{0}\,=\,\epsilon_{0}\,\,,
\label{cien}
\eeq
with $\Gamma_{\underline {12345}}$ being the product of the
$\Gamma$-matrices along the parallel coordinates
$x_{\parallel}^{i}$ ($i=\,1,\cdots, 5$). We are taking the convention
in which 
$\Gamma_{11}\,=\,\Gamma_{\underline 0}\,\Gamma_{\underline {12345}}\,
\Gamma_{\underline {\theta^1\theta^2\theta^3}}\,
\Gamma_{\underline{r}}$, in agreement with the array (\ref{oocho}). 
The $r$-dependent angle $\alpha_{(p,q)}(r)$ is given by:
\bear
\sin\,(\alpha_{(p,q)}(r))\,&=&\,
{p\,[\, H_{(p,q)}(r)\,]^{{1\over 2}}\over
\Big[\,p^2\,H_{(p,q)}(r)\,+\,g_s^2\,(\,q+\,p\chi_0)^2\,\Big]^{1\over
2}}\,\,,\rc\rc\rc
\cos\,(\alpha_{(p,q)}(r))\,&=&\,-
{g_s\,(\,q+\,p\chi_0)\over
\Big[\,p^2\,H_{(p,q)}(r)\,+\,g_s^2\,(\,q+\,p\chi_0)^2\,\Big]^{1\over
2}}\,\,.\rc
\label{ctuno}
\eear
Notice that when $(p,q)=(1,0), (0,1)$ and $\chi_0=0$ (\ie\ for the
NS5- and D5-brane, respectively), the angle defined by (\ref{ctuno})
is independent of $r$ and, in those cases, the Killing spinors only
depend on $r$ through the power of  the harmonic function in 
(\ref{nnueve}). In general, the dependence of $\epsilon_{(p,q)}$ on
$r$ will be given by the first two terms on the right-hand side of 
eq. (\ref{nnueve}) and will not be proportional to the unit matrix.
As we will verify in a moment, this fact will be crucial in our
analysis. Actually, by commuting the matrix appearing on the left-hand
side of eq. (\ref{cien}) with that of the right-hand side of 
(\ref{nnueve}),  one can obtain the following condition for 
$\epsilon_{(p,q)}$:
\beq
\Gamma_{\underline 0}\,\Gamma_{\underline {12345}}\,
\sigma_1\,\epsilon_{(p,q)}\,=\,
e^{i\alpha_{(p,q)}(r)\,\sigma_2}\,\,\epsilon_{(p,q)}\,\,.
\label{ctdos}
\eeq
By using the fact that $\epsilon_{(p,q)}$ has positive chirality, it
is easy to prove that eq. (\ref{ctdos}) is equivalent to:
\beq
\Gamma_{\underline{ 0 r}}\,\epsilon_{(p,q)}\,=\,
\,\sigma_{(p,q)}(r)\,
\Gamma_{\underline {0 \theta^1\theta^2\theta^3}}\,(i\sigma_2)\,
\epsilon_{(p,q)}\,\,,
\label{cttres}
\eeq
where $\sigma_{(p,q)}(r)$ is the following $r$-dependent linear
combination of the Pauli matrices $\sigma_3$ and $\sigma_1$:
\beq
\sigma_{(p,q)}(r)\,\equiv\,
\cos\,[\alpha_{(p,q)}(r)]\,\sigma_3\,+
\,\sin\,[\alpha_{(p,q)}(r)]\,\sigma_1\,\,.
\label{ctcuatro}
\eeq
In order to solve the supersymmetry preserving equation
(\ref{nocho}),  we must impose some additional constraint to the spinor 
 $\epsilon_{(p,q)}$. This extra condition is responsible of the fact
that our world-volume solitons preserve $1/4$ supersymmetry, instead
of the $1/2$ supersymmetry of the background. The  extra
requirement  we shall demand to $\epsilon_{(p,q)}$ is the one  which
follows from the fact that we have a D3-brane placed on the $(p,q)$
five-brane background. This supersymmetry condition associated to the
D3-brane probe has to be imposed locally \cite{kappa}, \ie\ at a
particular point of its world-volume. We shall choose this point to be
the one with polar coordinate $\theta^3=\theta=0$. Thus, we require
that:
\beq
\Gamma_{\underline {0 \theta^1\theta^2\theta^3}}\,(i\sigma_2)\,
\epsilon_{(p,q)}\Big|_{\theta=0}\,=\,\epsilon_{(p,q)}\Big|_{\theta=0}
\,\,.
\label{ctcinco}
\eeq
Using this condition, it is a simple exercise to find, by using eq. 
(\ref{nnueve}), the result of acting with $\Gamma_{\underline {0
\theta^1\theta^2\theta^3}}(i\sigma_2)$ on $\epsilon_{(p,q)}$ at
arbitrary angles:
\beq
\Gamma_{\underline {0 \theta^1\theta^2\theta^3}}\,(i\sigma_2)\,
\epsilon_{(p,q)}\,=\,e^{\theta\Gamma_{\underline{\theta r}}}\,
\epsilon_{(p,q)}\,=\,(\cos\theta\,+\,
\sin\theta\Gamma_{\underline{\theta r}}
\,)\,\epsilon_{(p,q)}\,\,.
\label{ctseis}
\eeq
By using eq. (\ref{ctseis}) in eq. (\ref{cttres}), one gets the result:
\beq
\Gamma_{\underline {0 r}}\,\epsilon_{(p,q)}\,=\,\sigma_{(p,q)}(r)\,
(\cos\theta\,+\,\sin\theta\Gamma_{\underline{\theta r}}
\,)\,\epsilon_{(p,q)}\,\,,
\label{ctsiete}
\eeq
which corresponds to the supersymmetry breaking condition required to a
$(-q,p)$-string in the radial direction at $\theta=0$. In order to
check this fact, one can verify for the NS5- and D5-brane backgrounds
that, indeed,  eq.  (\ref{ctsiete}) implies that the radial
spikes in these two cases  can be interpreted as a D1-brane and an
antifundamental string respectively. As a consequence of eqs.
(\ref{ctseis}) and  (\ref{ctsiete}), one can obtain that:
\bear
\Gamma_{\underline {0 \theta}}\,
\epsilon_{(p,q)}\,&=&\,\sigma_{(p,q)}(r)\,
(-\sin\theta\,+\,\cos\theta\Gamma_{\underline{\theta r}}
\,)\,\epsilon_{(p,q)}\,\,,\rc\rc
\Gamma_{\underline {\theta^1 \theta^2}}\,\epsilon_{(p,q)}\,&=&\,
(-i\sigma_2)\,\sigma_{(p,q)}(r)\,\Gamma_{\underline{\theta r}}
\,\epsilon_{(p,q)}\,\,.
\label{ctocho}
\eear
Eqs. (\ref{ctsiete}) and (\ref{ctocho}) allow us to evaluate the
right-hand side of eq. (\ref{nocho}). After performing this
evaluation, one gets two types of terms. First of all, one has terms
which contain the matrix $\Gamma_{\underline{\theta r}}$ which should
vanish. This condition leads to the equation:
\beq
\sigma_3\,\sigma_{(p,q)}(r)\,(\,r\sin\theta\,)'\,f_{\theta^1\theta^2}\,
-\,\sigma_1\,\sigma_{(p,q)}(r)\,f_{0\theta}\,-\,i\sigma_2\,
(\,r\cos\theta\,)'\,=\,0\,\,.
\label{ctnueve}
\eeq
We have, in addition, terms without $\Gamma_{\underline{\theta r}}$,
which give rise to the equation:
\bear
&&(\,r\sin\theta\,)'\,+\,\sigma_1\,\sigma_{(p,q)}(r)\,
(\,r\cos\theta\,)'
f_{\theta^1\theta^2}\,\,+\,i\sigma_2\,f_{0\theta}\,
f_{\theta^1\theta^2}\,=\,\rc\rc
&&=\,
\sqrt{(\,r^2\,+\,r'^2\,-\, f_{0\theta}^2\,)\,
(1\,+\,f_{\theta^1\theta^2}^2\,)}\,\,.
\label{ctdiez}
\eear
Eq. (\ref{ctnueve}) can be used to determine $f_{0\theta}$ and 
$f_{\theta^1\theta^2}$. Indeed, the left-hand side of this equation
contains terms proportional to the Pauli matrix $\sigma_2$, together
with others which are multiple of the unit matrix. By requiring these
two classes of terms to vanish independently, one gets two equations
which can be solved for $f_{0\theta}$ and 
$f_{\theta^1\theta^2}$ with the result:
\bear
f_{0\theta}\,&=&\,\cos[\alpha_{(p,q)}(r)]\,\,
(\,r\cos\theta\,)'\,\,,\rc\rc
f_{\theta^1\theta^2}\,&=&\sin[\alpha_{(p,q)}(r)]\,\,
{(\,r\cos\theta\,)'\over (\,r\sin\theta\,)'}\,\,.
\label{ctonce}
\eear
Quite remarkably, one can verify that the values of 
$f_{0\theta}$ and $f_{\theta^1\theta^2}$ given in eq. (\ref{ctonce})
also satisfy eq. (\ref{ctdiez}) and, thus, we have succeeded to find
the world-volume field strengths corresponding to the D3-brane
supersymmetry condition (\ref{ctcinco}). It remains to verify that the
values we have just found are the same as those obtained in section 4 
(eqs. (\ref{setenta}) and (\ref{sttres})), by means of the energy
bound argument. The first step in this verification will consist in
rewriting the values of $\sin[\alpha_{(p,q)}(r)]$ and
$\cos[\alpha_{(p,q)}(r)]$ in terms of the dilaton field in the 
$(p,q)$ five-brane background. By combining eqs. (\ref{nueve}) and
(\ref{ctuno}), one arrives at:
\bear
\sin[\alpha_{(p,q)}(r)]&=&p\,e^{-{\phi\over 2}}\,\,
{[\, H_{(p,q)}(r)\,]^{{1\over 4}}\over 
[\, \rho_{(p,q)}\,]^{{1\over 2}}}\,\,,\rc\rc\rc
\cos[\alpha_{(p,q)}(r)]\,&=&\,-\,g_s\,(\,q\,+\,p\,\chi_0\,)\,
e^{-{\phi\over 2}}\,\,
{[\, H_{(p,q)}(r)\,]^{-{1\over 4}}\over 
[\, \rho_{(p,q)}\,]^{{1\over 2}}}\,\,.
\label{ctdoce}
\eear
By substituting  these results on the right-hand side of eq.
(\ref{ctonce}) and, after using the definitions of 
$f_{0\theta}$ and $f_{\theta^1\theta^2}$ (eq. (\ref{nsiete})), it is
straightforward to check that the values of $F_{0\theta}$ and 
${\cal F}_{12}$ obtained from (\ref{ctonce}) coincide with the ones
displayed in eqs. (\ref{setenta}) and (\ref{sttres}) (recall that we
are adopting the gauge  (\ref{ocho}) for the NS field $B$). Therefore,
we conclude that the BPS conditions we have used to minimize the
energy of the D3-brane probe are, indeed, the ones needed to preserve
$1/4$ supersymmetry. 

\bigskip
\setcounter{equation}{0}
\section{Concluding remarks}
\medskip

In this paper we have studied the world-volume solitons of a D3-brane
probe in the background of a stack of parallel $(p,q)$ five-branes. We
have used $S$-duality to find an energy bound,  whose saturation is
achieved when a certain first-order BPS equation is satisfied. By
solving this BPS equation we have determined the D3-brane embeddings
in the $(p,q)$ five-brane geometry. The resulting configurations of the
brane probe contain spikes, which can be interpreted as a
bundle of $(-q,p)$ strings emanating from the D3-brane world-volume.

An important point in  our analysis is the r\^ole played by the
background which, through its coupling to the D3-brane in the
WZ term of the action $S_{WZ}$, provides a source for the world-volume
gauge field. For the background we studied two different terms in
$S_{WZ}$ are relevant. Moreover, we have verified that the
configurations which saturate the energy bound are precisely those
which are $1/4$ supersymmetric. This result is very important since 
it provides a precise interpretation of the BPS equation obtained by
the energy argument. 

It is clear that the ansatz we have adopted to minimize the energy is
not the most general one. We could take, for example, a radial
deformation of the brane probe with less symmetry than the $SO(3)$
invariant one we have considered here. In these new configurations the
coordinate $r$ would depend on the three angles $\theta^i$ and, in
principle, one could apply  the methods of ref. \cite{kappa} to study
them. The resulting BPS equation for this case would be more complicated
although, from the results of \cite{kappa}, one expects to find at
least  some of its particular solutions. 

It would be also very interesting to
apply the methodology followed here to other background and probes. In
particular we could apply different chains of dualities (including
T-duality, which was not used here) to generate new world-volume
solitons from the known ones.

\section{ Acknowledgments}
We are grateful to J. M. Camino, D. Mateos, J. Gomis and J. Sim\'on
for very useful discussions.  We thank I. P. Ennes for a critical
reading of the manuscript. This work was supported in part by DGICYT
under grant PB96-0960,  by CICYT under grant  AEN96-1673 and by the
European Union TMR grant ERBFMRXCT960012.

\end{document}